\documentclass{aa}
\usepackage{graphicx}
\usepackage{txfonts}
\usepackage{color}
\def\ltapprox{\raise 2pt \hbox {$<$} \kern-1.1em \lower 5pt \hbox {$\approx$}}

\def\ltsim{\; \raise0.3ex\hbox{$<$\kern-0.75em \raise-1.1ex\hbox{$\sim$}}\; }
\def\gtsim{\; \raise0.3ex\hbox{$>$\kern-0.75em \raise-1.1ex\hbox{$\sim$}}\; }

\def\arcmin{$^{\prime}\,$}

\begin{document}

\title{Is the Sunyaev-Zeldovich effect responsible for the 
observed steepening in the spectrum of the Coma radio halo ?}

\author{G. Brunetti\inst{1}\fnmsep\thanks{\email{brunetti@ira.inaf.it}},
L. Rudnick\inst{2}, R. Cassano\inst{1}, P. Mazzotta\inst{3}, 
J.Donnert\inst{1}, K. Dolag\inst{4}} 
\authorrunning{G. Brunetti et al.}
\titlerunning{Coma spectral steepening}

   \offprints{G.Brunetti}

   \institute{INAF - Istituto di Radioastronomia, via P. Gobetti 101,
   I-40129 Bologna, Italy\\
	\and
   Minnesota Inst. for Astrophysics, School of Physics \&
	Astronomy, Univ. of Minnesota, 116 Church Street SE, Minneapolis, MN
	55455, USA\\
	\and 
   Dipartimento di Fisica, Universita' degli Studi di Roma ``Tor
	Vergata,'' via della Ricerca Scientifica, 1, I-00133 Roma, Italy\\
	\and
   Universit\"atssternwarte M\"unchen, Scheinerstr. 1, D-81679
   München, Germany}

   \date{Received...; accepted...}

\abstract{}{
The radio halo in the Coma cluster is unique in that its
spectrum has been measured over almost two decades in frequency.
The current radio data show a steepening of the spectrum at higher
frequencies, which has implications for models of the radio halo origin. 
There is an on-going debate on the possibility that 
the observed steepening of the spectrum and the 
apparent shrinking of the 
halo-size at higher frequencies is not intrinsic to the emitted
radiation, but 
is instead caused by 
the Sunyaev-Zeldovich (SZ) effect.}
{Recently, the Planck satellite obtained unprecedented measurements of the 
SZ signal and its spatial distribution in the Coma cluster, 
allowing a conclusive testing of this hypothesis.
Using the Planck results, we calculate the modification of the 
radio halo spectrum by the SZ effect in three different ways.
With the first two methods 
we measured the SZ-decrement by adopting self-consistently the aperture radii 
used for flux measurements of the radio halo at the different frequencies.
First we adopted the global compilation of data-points from Thierbach et
al. and a reference aperture radius consistent with those used by 
various authors.
Second we used the available brightness profiles of the halo at different
frequencies to derive the spectrum of the halo within two fixed
apertures, corresponding to the size of the halo measured 
at 2.675 and at 4.85 GHz, 
and derived the SZ-decrement using these apertures.
As a third method we used the quasi--linear 
correlation between the $y$--signal and
the radio-halo brightness at 330 MHz 
discovered by the Planck collaboration to derive 
the modification of the synchrotron spectrum 
by the SZ-decrement in a way that is almost independent of the adopted 
aperture radius.}
{ We found that the spectral modification induced 
by the SZ-decrement is 4--5 times smaller 
than those necessary to explain the observed steepening 
at higher frequencies.
We also show that, if a spectral steepening is absent from the
emitted spectrum, 
future deep observations at 5~GHz with
single dishes are expected to measure a halo flux in a 40 arcmin
aperture-radius that would be $\sim$7-8 times higher than currently
seen, thus providing a complementary test to our findings.}
{We conclude that the current radio data of the Coma radio halo
suggest the presence of a break or cut-off in the spectrum of the emitting 
electrons at energies of a few GeV.}


\maketitle

\section{Introduction}

Giant radio halos are diffuse synchrotron radio sources of Mpc-scale in
galaxy clusters. They are observed in about $1/3$ of the X--ray 
luminous galaxy clusters (e.g. Giovannini et al. 1999; 
Kempner \& Sarazin 2001; Cassano et al. 2008; Venturi et al. 2008), 
in a clear connection
with dynamically disturbed systems (Buote 2001,
Govoni et al. 2004, Cassano et al. 2010, 2013).
The connection between cluster mergers and radio halos suggests that
these sources trace the hierarchical cluster assembly and probe the
dissipation of gravitational energy during the dark-matter-driven
mergers that lead to the formation of clusters.
However,  the details of the
physical mechanisms responsible for the generation of synchrotron
halos are still unclear.

\noindent
Two main scenarios are advanced for the origin of these sources. One,
the ``reacceleration'' model, is based on the idea that seed relativistic
electrons are re-accelerated by turbulence produced during merger events
(Brunetti et al. 2001; Petrosian 2001; Fujita et al 2003; Cassano \&
Brunetti 2005; Brunetti \& Lazarian 2007;
Beresnyak et al. 2013). The alternative is that cosmic
ray electrons (CRe) are injected 
by inelastic collisions between long-lived relativistic protons (CRp)
and thermal
proton-targets in the intra-cluster-medium (ICM) 
(the ``hadronic'' model, Dennison 1980; Blasi \&
Colafrancesco 1999; Pfrommer \& En\ss lin 2004, PF04; Keshet \& Loeb 2010). 
More general calculations attempt to combine the two mechanisms 
by modeling the reacceleration of relativistic protons and their secondary
electrons (e.g. Brunetti \& Blasi 2005; Brunetti \& Lazarian 2011).

Concerns with a purely hadronic origin of radio halos arise from
the large energy content of CRp that is necessary to explain
radio halos with very steep spectra (Brunetti 2004; PE04; Brunetti et
al 2008; Macario et al 2010) and from the non-detection of
galaxy clusters with radio halos in the 
$\gamma$--rays (Ackermann et al. 2010;
Jeltema \& Profumo 2011; Brunetti et al. 2012).

Turbulent reacceleration of CRe and the production of secondary cosmic
rays via hadronic collisions leave different imprints 
in the spectra of the cluster non-thermal emission.
In re-acceleration
models,  the radio spectra are determined by the low efficiencies
of the acceleration mechanism, allowing
the acceleration of CRe only up to energies of several GeV, where
radiative, synchrotron and inverse Compton (IC),  
losses become stronger and quench the
acceleration process (Schlickeiser et al. 1987; Brunetti et al. 2001; 
Petrosian 2001). This effect leads to spectra that may steepen at high radio
frequencies and a variety of spectral shapes and slopes. 
Pure hadronic models, by
contrast, have spectra with fairly smooth power-laws extending to very
high frequencies. Significant spectral breaks at radio frequencies would, 
in the
hadronic model,  imply an 
unnatural strong break in the spectrum of the primary CRp at energies
10--100\,GeV (eg. Blasi 2001). 
Although the spectra of radio halos are difficult to measure, with only
a handful of good-quality data-sets available to date, 
studying them provides one of the most promising 
ways to shed light on the origin of these sources 
(e.g., Orru' et al. 2007; 
Kale \& Dwarakanath 2010; van Weeren et al. 2012; Venturi et al. 2013; 
Macario et al 2013), 
especially in view of the new generation of low--frequency radio 
telescopes, such as the Low Frequency Array (LOFAR) and 
the Murchison Widefield Array (MWA).

\noindent
The radio halo in the Coma cluster
is unique in that its spectrum has been measured over almost two decades
in frequency  (Fig.~1).
The observed steepening at high frequencies was used to argue
against a hadronic origin of the halo
(Schlickeiser et al. 1987; Blasi 2001; Brunetti et al. 2001; Petrosian 2001).
En\ss lin (2002, E02) first proposed that such a steepening
may be caused (at least in part) by the thermal Sunyaev-Zeldovich (SZ)
decrement seen in the
direction of the cluster. This possibility was further elaborated by
PE04, concluding that a simple 
synchrotron power-law spectrum, with spectral index 
$\alpha = 1.05-1.25$, provides a fair description of the observed radio
spectrum after taking into account the negative flux bowl due to
the SZ effect (see also En\ss lin et al 2011).
Other efforts to model the SZ-effect in the Coma cluster, however,
concluded that this effect is not significant (Reimer et al 2004;
Donnert et al 2010).
All these attempts used the $\beta$--model spatial distribution of the
ICM from X--ray observations\footnote{Donnert et al.~used the Coma cluster
from a constrained cosmological simulation of the local Universe.}, but the
differences between the two lines of thought comes essentially from the
apertures used to estimate the SZ-decrement.

\noindent
The Planck satellite has recently obtained resolved and
precise measurements of the SZ signal in the Coma cluster 
(Planck Collaboration X (2012), PIPX).
These data strongly reduce 
the degrees of freedom and uncertainties in the modeling of 
the SZ contribution
and allow a straightforward test of the spectral steepening hypothesis. 
This is the aim of the present paper.

\noindent
A $\Lambda$CDM cosmology ($H_{o}=70\,\rm km\,\rm s^{-1}\,\rm Mpc^{-1}$, 
$\Omega_{m}=0.3$, $\Omega_{\Lambda}=0.7$) is adopted.

\section{Thermal Sunyaev-Zeldovich effect}

The inverse Compton interaction of photons of the Cosmic Microwave
Background (CMB) with hot
electrons in the ICM modifies the CMB spectrum, with 
respect to a blackbody, by a quantity $\delta F_{SZ}$ that
depends on frequency (see e.g., Carlstrom et al. 2002 for a review).
For low optical depths this quantity measured on an 
aperture $\Omega$ is 

\begin{equation}
\delta F_{SZ}(\nu,\Omega) = 2 {{(k_B T_{cmb})^3}\over{(h c)^{2}}} 
f(\nu) \int_{\Omega} y d\Omega^{\prime} \, ,
\end{equation}

\noindent
where the cluster Compton parameter is 

\begin{equation}
y = {{\sigma_T}\over{m_e c^2}}
\int P_e(r) dl \, ,
\end{equation}

\noindent
$l$ is the line of sight,  
and the spectral distortion at the frequency $\nu$ is 

\begin{equation}
f(\nu) = {{x_{\nu}^4 e^{x_{\nu}}}\over{(e^{x_{\nu}} -1)^2}}
\left[ {x_{\nu} \over{\tanh(x_{\nu}/2)}} -4 \right] 
\stackrel{x_{\nu} \rightarrow 0}{\longrightarrow}
-2 x_{\nu}^2 \, ,
\end{equation}

\noindent
where $x_{\nu}= {\rm h} \nu/k_B T_{cmb}$. 
At GHz frequencies $x_{\nu} << 1$, and 
the SZ effect from the ICM in galaxy clusters
creates a negative flux bowl on the scale of the clusters.
The flux from radio sources measured by single-dish radio telescopes is
that in excess of  the background ``zero'' level that is determined on 
larger scales. Once discrete sources 
are subtracted appropriately, the resulting flux observed from radio
halos in an aperture $\Omega_H$ is $F_{obs}(\nu, \Omega_H) = 
F(\nu, \Omega_H) + \delta F_{SZ}(\nu, \Omega_H)$, $F$ is relative
to the intrinsic emission.
The SZ-decrement (negative) signal, $\delta F_{SZ}$, will become
comparable to the flux from the halo at sufficiently high frequencies, 
leading to an apparent steepening of the observed spectrum
(Liang et al 2000, E02).

\begin{figure}
\begin{center}
\includegraphics[width=0.425\textwidth]{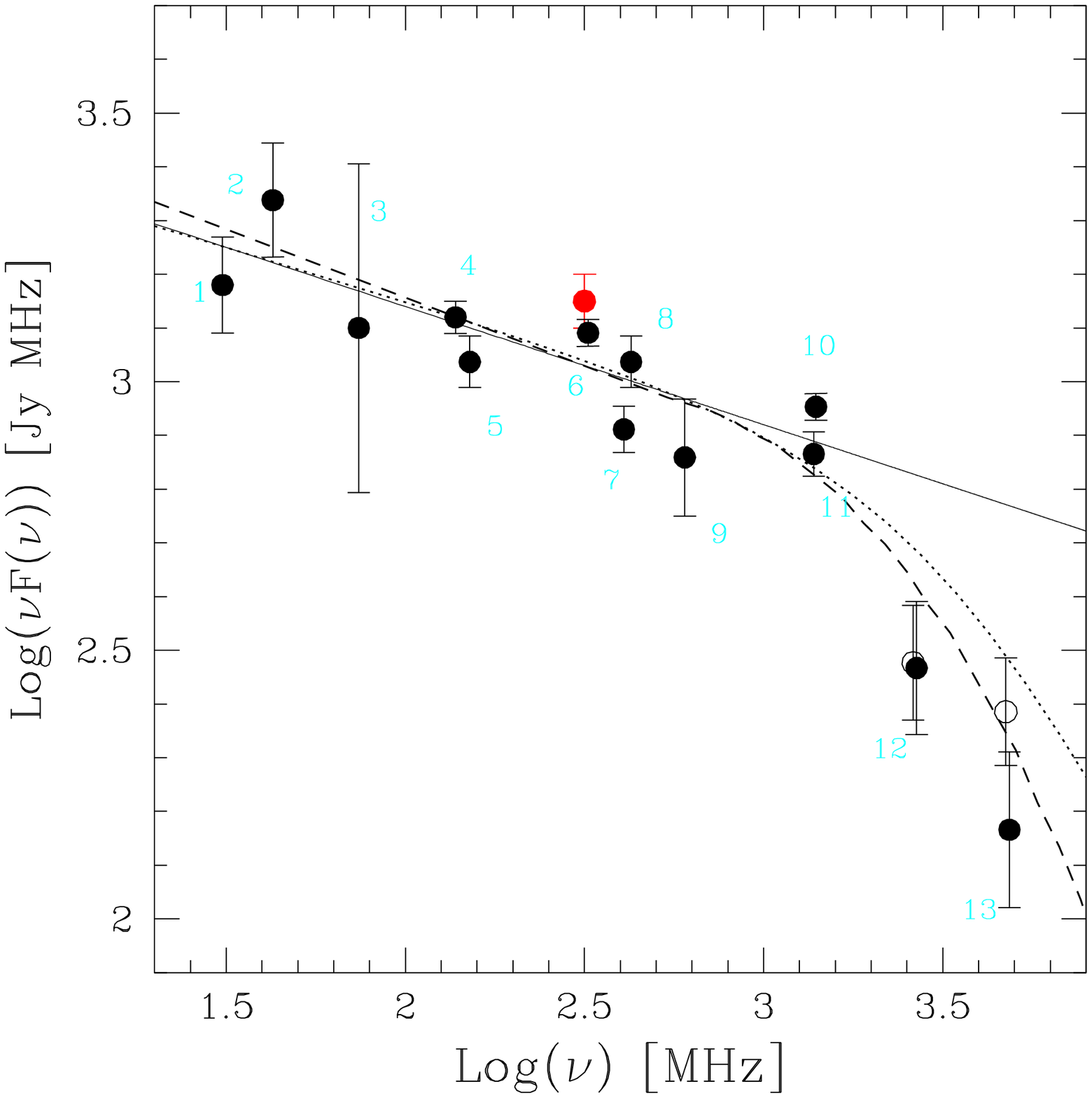}
\caption[]{Observed
spectrum of the Coma radio halo (black data-points) and the 
power law ($\alpha = 1.22 \pm 0.04$) that best fits the
spectrum at {\it lower} frequencies, $\nu \leq 1.4$ GHz (solid line).
Empty points are the data with the SZ correction added, the SZ-decrement
is calculated from Planck measurements by adopting an aperture radius
$= 0.48 R_{500}$.
The dotted line is a synchrotron model assuming a broken power-law energy
distribution of the emitting electrons, 
$N(E) \propto E^{-\delta}$ and $\propto E^{-(\delta+\Delta \delta)}$, at
lower and higher energies, with $\delta =2.4$ and
$\Delta \delta = 1.6$.
The dashed line is a synchrotron model assuming a power-law ($\delta =2.5$)
with a high energy cut-off, which occurs for instance in 
(homogeneous) reacceleration models (see eg. Schlickeiser et al.~1987).
The red point is the flux measured in the high-sensitivity observations
at 330 MHz by Brown \& Rudnick (2011) within
an aperture radius =$0.48 R_{500}$.}
\label{Fig.Lr_RH}
\end{center}
\end{figure}

\section{Implications for the Coma radio halo}

In this section we use Planck measurements to calculate 
the modification of the
spectrum of the Coma radio halo at higher frequencies
that is caused by the SZ-decrement.  Because this correction depends
on the aperture over which it is calculated, as well as no the varying radio 
sensitivities
and observed halo sizes in the literature, we used three 
different approaches to ensure that our results are robust.  

\begin{itemize}

\item (i) We used the halo spectrum from the data compilation from
Thierbach et al.(2003, T03) and calculated the SZ decrement in a 
reference aperture radius that is
consistent with those used at the various frequencies in the T03 compilation;

\item (ii) we used the subset of radio observations for which
the actual radial
profiles are available to derive the halo spectrum 
in two fixed aperture radii and self-consistently calculated 
the SZ decrement within these radii;

\item (iii) we used the correlation between 
radio brightness and y-parameter discovered by PIPX to derive the ratio of 
radio--halo flux and negative SZ flux as a function of frequency and
almost independently of the aperture radius.
\end{itemize}

\begin{figure*}
\begin{center}
\includegraphics[width=0.95\textwidth]{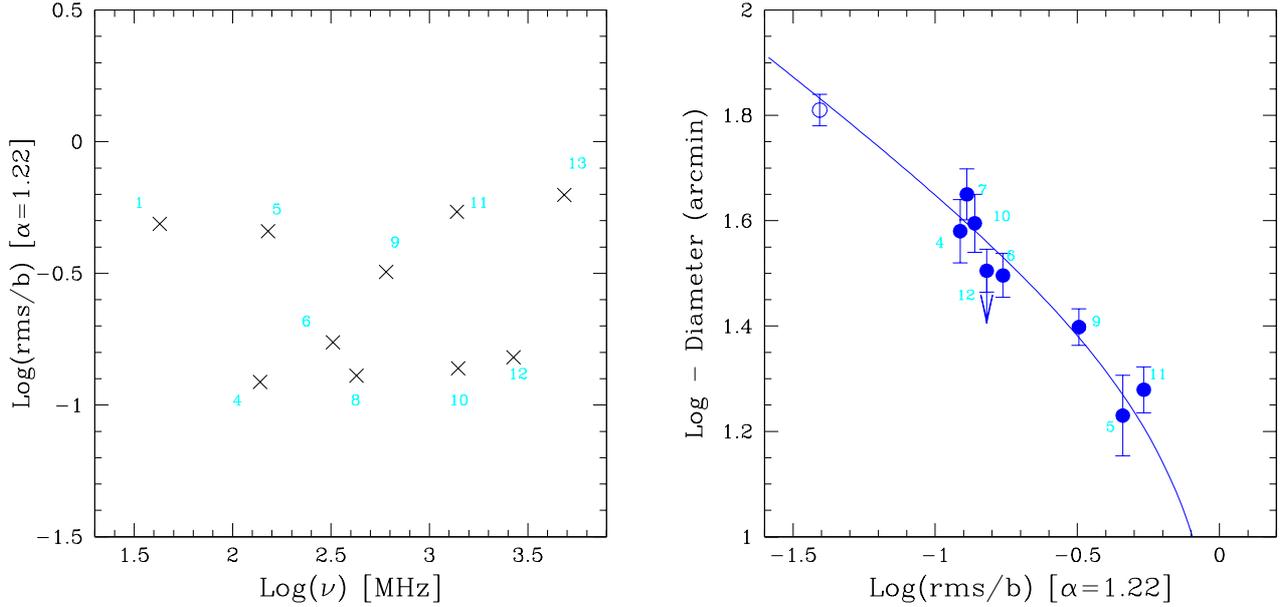}
\caption[]{(Left): sensitivity to diffuse emission of the observations at
different frequencies  (rms$/$beam, scaled
to 0.3~GHz with $\alpha=1.22$, using an arbitrary scale). 
(Right): radio halo diameters measured from radio maps (using $3
\sigma$ contours) vs sensitivity of the observations; for comparison, the
empty symbol marks data from
the Brown \& Rudnick (2011) WSRT observation at 330 MHz.
For the Effelsberg data-point at 2.675 GHz (12) we used an upper limit 
because the size of the halo measured from the contours of the image
by T03 is biased high by discrete sources embedded
in the halo.
}
\label{Fig.Lr_Lx}
\end{center}
\end{figure*}

\noindent
Before using these different methods for evaluating the SZ contribution,
we first discuss the main uncertainties in the radio halo spectrum,
in particular 
the scale size of the halo which is essential for the
application of these methods.

\noindent
Fig.~1 shows the observed spectrum from
the T03 compilation of data--points\footnote{we also add WSRT
data at 139 MHz (Pizzo 2010)} and the best fit to the data 
at lower ($\leq 1.4$ GHz)
frequencies, a power law with $\alpha = 1.22 \pm 0.04$.
We note that the scatter of the data is larger than expected from the
quoted errors (the best fit has
$\chi^2_{\rm red}=2.65$), which suggests 
systematics that are likely due to the different sensitivities 
and the variety of telescope systems used over the past 30 years.
To investigate the effects of sensitivity on the observed halo sizes 
and fluxes  
we returned to the original papers used for Fig.~1.  
Wherever the information was available, 
we show in Fig.~2 (left) the
sensitivities to diffuse emission of the observations at the
different frequencies (sensitivities  are all scaled 
to 0.3~GHz using $\alpha =1.22$).
If we focus on the data at $\nu \leq 1.4$ GHz (points 1-11), the comparison 
between
Fig.~1 and 2 (left) immediately shows that fluxes from 
higher-sensitivity observations (points 4,6,8,10) are systematically biased 
high (about 50\% higher)
with respect to those derived from the less sensitive observations
(points 1, 5, 9, and 11).
This is because 
better sensitivities allow one to trace the diffuse emission of the
halo to larger
distances from the cluster center. This is clear from Fig.~2 (right), 
where we show the diameters of the radio halo that we measured
from the 3$\sigma$ contours in the published radio maps 
as a function of the sensitivity. 
The measured 3$\sigma$ diameter ($=2 \sqrt{R_{min} R_{max}}$, $R_{min}$
and $R_{max}$ are the smallest and largest radii in the map)
increases with sensitivity in a way that
depends on the brightness distribution of the radio halo:
in Fig.~2 (right) we also show the behavior expected assuming a 
radio-brightness
distribution of the form $I_R 
\propto (1 + (r/r_c)^2)^{-k}$, where $r_c$ is the X-ray core radius
of Coma, $r_c = 10.5$ arcmin, and where $k \simeq 0.7*(3\beta -1/2)$, as
inferred by Govoni et al. (2001) ($\beta = 0.75$).  
This provides a good description of the data;  
note that points 5 and 11, with
the poorest sensitivities, are both approximated well by the model, 
but come from very different frequencies, 151~MHz and 1.38~GHz, 
respectively.

\noindent
With a good estimate of the sensitivity dependence, we can now
address a key concern, i.e., whether the observed spectral 
steepening is caused by a lack of sensitivity.  For point 12,
at 2.675~GHz , the sensitivity is the same as for a number of points 
from 139~MHz to 1.4~GHz, therefore it is expected to appear on the
power-law line (Fig.~1) if there is no spectral steepening.  
However, it is a factor of $\sim$2 below this line, supporting the case
for actual spectral steepening, whether intrinsic or caused by the SZ effect.
We also note that the flux of the Coma radio relic measured
at 2.675 GHz by T03 is indeed consistent with the power-law shape derived from other
observations in the range 151 MHz--4.75 GHz (T03, Fig.~8), adding confidence
on the steepening measured by T03 for the halo.
The effect of sensitivity on the highest frequency point, 13, can be best 
assessed by comparison with points 1, 5 and 11, from 30~MHz, 151~MHz
and 1380~MHz, respectively, which have a similar sensitivity.
The flux of points 1, 5 and 11 is $\sim$20\% lower than the 
power-law line in Fig.1.    
A drop of 20\% for point 13 does not explain its flux, which is more
than a factor of $\sim$3 below the power-law line.  
We conclude that the spectral steepening at 2.675~GHz and 4.85~GHz
is not an observational effect caused by the sensitivity of the
observations at the different frequencies.

Systematics in the spectrum might also come from
the different procedures used to subtract discrete sources in the halo region. 
The subtraction becomes critical at higher frequencies where most of the flux
in the halo region is associated with  discrete sources.
The flux at 4.85 GHz of the two brightest sources in the halo region 
(NGC 4869 and 4874) is known from VLA observations
(175 mJy in total, see T03
and Kim 1994),
while additional 55 mJy are attributed by T03 to discrete sources
in the halo region by using the master--source list of Kim (1994),
which 
also reports spectral indices.
Most of the spectral information in Kim (1994) 
was obtained at frequencies $\leq 1.6$ GHz, therefore the flux of discrete 
sources may be overestimated if their spectral indices actually steepen
at higher frequencies (T03); this would bias the resulting
flux of the halo low.
We investigated this effect by assuming 
the typical spectral steepenings between low and high (1.4
-- 4.8 GHz) frequencies that are measured for samples of 
radio sources, $\Delta \alpha \leq 0.15$ (Kuehr et al. 1981;  
Helmboldt et al. 2008), and found that 
the increment of the radio halo flux with respect to that from 
T03 is $< 30$\%, which is lower than the error reported in Fig.~1.
Again we conclude that the spectral steepening observed at high
frequencies is not driven by obvious observational biases.

As a final step we studied the possible effect caused by the 
different adopted flux calibration
scales, because spectral measurements were taken over a long time-span.
The best modern values are presented by Perley \& Butler (2013).  
Using their values for the calibrator
3C286, we corrected the T03 2675 
and 4850 MHz data.  This resulted in a reduction of their fluxes by 3.5\% 
and 2.1\%, enhancing the steepening by a tiny amount, well within the 
errors.
Most of the other papers summarized by T03 do not explicitly 
describe their flux scale, although the most common scale in usage was 
that of Baars et al. (1977).
Down to 326~MHz, which is the lowest frequency studied by Perley \& 
Butler, the other literature values would decrease by an average of 
1.6\%, with the largest decrease being 2.5\% for the 608.5~MHz point, 
producing no significant change in the spectrum\footnote{
When we evaluated the corrections to the halo spectrum below
300 MHz by extrapolating the Perley \& Butler analytic fit to the 
spectrum of 3C 286, we
found a marginal increment of the halo flux below 50 MHz.}.

\noindent
We proceed to examine the potential contribution of the SZ effect.

\subsection{Method (i)}

Our first method of evaluating the SZ contribution involves 
using the fluxes as reported in the literature and adopting an
appropriate ``reference'' aperture radius for the correction.  
Fig.~3 (left) shows the ``equivalent diameter'' (defined below) of the
regions that is used by the respective authors for flux measurements, 
wherever these are available.  We deliberately biased 
our calculations toward higher SZ contributions by using 
the (larger) apertures at  $\nu \leq 1.4$ GHz; at these low frequencies, 
the SZ decrement is negligible and cannot artificially make the halo 
look smaller.  
In some cases the fluxes were
taken from boxes or from complex (non circular) regions 
that encompass the scale where diffuse emission is detected,
therefore we define the ``equivalent diameter'' as $2 \sqrt{A / \pi}$, 
$A$ being the area from
which fluxes in the literature were extracted.

\noindent In Fig.~3 we see that for the high-sensitivity points
4, 6, and 10 (139, 330, and 1400~MHz),  the aperture radii are
consistent 
with a value of 23\arcmin ($=0.48 R_{500}$ , where
$R_{500}=$47 arcmin$=1.3$ Mpc, PIPX).
The low-sensitivity point 1 at 30~MHz is consistent with this as well. 
These points with consistent aperture define the power law seen in
Figure~1; the low-sensitivity point 5, at 151 MHz, has a smaller size
and is indeed $\sim 20$\% below the power law in Fig.~1.
We therefore adopted 23\arcmin as the radius over which to calculate 
the SZ correction
\footnote{Here we ignore point 7, at 430~MHz (Hanisch 1980), which
shows a larger radius, to have a consistent 
aperture within which the fluxes are measured. However. including it 
has no effect on the spectral fit (Fig.~1).
In addition, we note that 
although its larger radius should have led to a higher flux, oversubtraction 
of point sources in the
central region of the halo, which led to a bowl in the Hanisch (1980) 
map (Fig.~1c in Hanisch 1980), resulted in fluxes consistent with the
23\arcmin power law.}.

With this 23\arcmin  ($0.48 R_{500}$ ) radius, we corrected 
for the SZ decrement on scales significantly larger than 
the observed halo sizes at the high frequencies (points 12, and 13), 
and thus we eventually overestimated the actual effect.   
From Planck measurements of the Compton parameter $y$ (PIPX) and
from Eqs.1--3, we found $\delta F_{SZ}= -(1.08 \pm 0.05)
(\nu / GHz)^2$ mJy. This is about four times lower than that used
by PE04 (their eq.~74) and about 30\% higher than
that used in Donnert et al. (2010).
The main reason for the discrepancy with PE04 
is that they calculated the SZ-signal by integrating Eq.1 over 
an excessively large aperture, $R = 5 h_{50}^{-1}$Mpc (in Eq.1 
$\Omega =
2\pi \int_0^{R/D_A} d\theta$, $D_A$ the angular distance).
This corresponds to
$\sim 2.75 R_{500}$, which indeed is 5-6 times higher than the
radius of the radio halo in the observations of the T03 compilation
(Fig.3 left)
\footnote{Within this aperture we found that 
the SZ-decrement measured by Planck is
$\simeq -3.7 (\nu/GHz)^2$mJy. This is similar, although slightly
smaller, than the
PE04 value, which was calculated using X-ray data,
assuming an isothermal (kT=8.2 keV) and spherical cluster
with the spatial distribution of the gas density given by the the
extrapolation
of the beta-model to $R = 5 h_{50}^{-1}$Mpc distances.}.
Fig.~1 shows the high-frequency data-points with the SZ
correction added (filled symbols).
We conclude that the SZ-decrement is not important:
an SZ-decrement about four times larger than that measured
by Planck would be needed to reconcile the data-points at higher frequency
with a power-law spectrum.

\noindent With this result, we must conclude that the spectral 
break above 1.4~GHz is intrinsic to the source.
We attempted to constrain the magnitude of the break by fitting 
the data-points in Fig.~1, corrected for
the SZ-decrement (adding a flux $= 1.08 (\nu / GHz)^2$ mJy),
with a broken power-law with slopes
at lower and higher frequencies $\alpha$ and $\alpha + \Delta \alpha$.
An F-test analysis, using the statistical errors from Fig.1, 
constrains $\Delta \alpha > 0.45$ (90\% confidence
level), implying a corresponding break
in the spectrum of the emitting electrons $\Delta \delta > 0.9$.
A strong break is also clear from Fig.~1, which shows synchrotron 
models assuming
both a break $\Delta \delta = 1.6$ (dotted line) and a high-energy
cut-off (dashed line). A strong break (or cut-off) in the spectrum
of the emitting electrons is thus required by the current data
even after correcting for the SZ effect.

\subsection{Method (ii)}

As a second approach, we directly used the brightness profiles 
of the radio halo at different frequencies wherever these were available. 
This avoided problems associated with measurements at better 
sensitivities, which can be integrated to larger radii.  
In particular, we constructed new spectra by integrating the lower-frequency 
fluxes only out to radii of 17.5\arcmin and 
13\arcmin (0.37 and 0.27 $R_{500}$), which correspond to the effective 
radii at 2.675 and 4.85~GHz, respectively (as in T03). 
The calculated SZ decrements from
the Planck observations are  $=-0.82 \pm 0.03$ 
and $-(0.50\pm 0.02)(\nu / GHz)^2$ mJy, respectively. 

\noindent In Fig.~4 we show the reconstructed spectrum (SZ-corrected) 
within these radii. 
As in method (i), we found that the SZ
decrement is negligible and that a correction 4-5 times
larger than that implied by Planck measurements would be needed to
reconcile the data at 2.675 and 4.85 GHz
with the best-fit spectrum obtained at lower frequencies.

\noindent We furthermore note that Fig.~4 shows indications for a steepening
of the halo spectrum with increasing aperture radius.
The best-fit slope obtained using the
smallest aperture $=$13\arcmin ($=0.27 R_{500}$),  
$\alpha = 1.10 \pm 0.01$, is smaller than that obtained 
with an aperture of 17.5\arcmin ($=0.37 R_{500}$), 
$\alpha = 1.17 \pm 0.02$, and than the best-fit slope to the TH03
compilation of data, which refer to larger apertures (Sect.~3.1).
This qualitatively agrees with the radial spectral
steepening of the Coma radio halo that was
reported by Giovannini et al.~(1993)
and Deiss et al.(1997).

\subsection{Method (iii)}

In a third approach we 
evaluated the importance of the SZ decrement 
by using the correlation found by PIPX
between the y-signal and the radio flux in a beam area,
$F(0.3, \Omega_b)$, from recent
deep WSRT observations at 330 MHz (Brown \& Rudnick 2011).
This is $y = 10^{-5} \times 10^{(0.86\pm 0.02)}
(F(0.3, \Omega_b)/Jy)^{0.92 \pm 0.04}$, using a
10-arcmin FWHM beam and measured to a maximum radial distance 
$\sim 0.8 \times$ $R_{500} \sim 38$ arcmin.
From Eq.~1, this point-to-point radio-SZ correlation
can be converted into a relation between the SZ-decrement integrated
over a beam area $\Omega_b$, $\delta F_{SZ} (\nu, \Omega_b) 
= 2(k_B T)^3f(\nu) y \Omega_b/(hc)^2$,
where $\Omega_b \simeq 9.6 \times 10^{-6}$rad$^2$ and
the radio flux integrated in the same beam.
Assuming that the radio halo has an intrinsic
power-law spectrum, $F(\nu) \propto \nu^{-\alpha}$, it is

\begin{equation}
{{ \delta F_{SZ}(\nu, \Omega_b) }\over{ F(\nu, \Omega_b) }}
\simeq -{{ 1.2 \times 10^{-4} }\over{ 0.33^{\alpha} }}
( {{\nu}\over{GHz}} )^{2 + \alpha}
\left( {{F(0.3, \Omega_b)}\over{Jy}} \right)^{-0.08 \pm 0.04}
\, .
\end{equation}

\noindent
Here the ratio $\delta F_{SZ} / F$ is calculated within a beam area 
($\Omega_b = 9.6 \times 10^{-6}$rad$^2$) and depends on the distance from
the cluster center because $F$, on the right-hand side of Eq.~4, 
decreases with distance. 
However, the quasi--linear scaling between $\delta F_{SZ}$
and $F$ makes this dependence very weak.
Indeed, according to Fig.~9 in PIPX, $F(0.3,
\Omega_b)$ ranges from $\sim 1$ Jy in the central halo regions
to $\sim 0.06$ Jy in the periphery, implying a maximum
variation of $<$25\% in the ratio $\delta F_{SZ} / F$ 
as a function of distance.
Such a weak dependence allows us to readily also derive a ratio
$\delta F_{SZ} / F$ referred to a larger aperture.
If we assume the average value of $F$ within the halo region,
$F \sim 0.2$ Jy (PIPX), we obtain the ratio $\delta F_{SZ} / F$ on
the aperture of the halo, $\Omega_H$ (with $R_H \sim 0.85 R_{500}$) 

\begin{equation}
{{ \delta F_{SZ}(\nu, \Omega_H) }\over{ F(\nu, \Omega_H) }}
\sim - 1.4 \times 10^{-4} {{(\nu/GHz)^{2 + \alpha} }\over{0.33^{\alpha} }}
\big{\langle} ({{F(0.3, \Omega_b)}\over{ 0.2 Jy}})^a 
\big{\rangle}_{\Omega_H}
\, ,
\end{equation}

\noindent
where $a=-0.08$ and 
$\langle .. \rangle$ is a flux-weighted average on the halo aperture. 

\noindent
The ratio $\delta F_{SZ} / F$ from Eq.~5 at different frequencies is
shown in Fig.~5 assuming different (intrinsic) spectral slopes.
Assuming a power law with slope $\alpha \sim 1.22$, the spectrum
of the halo at lower frequencies,
we found that the negative signal caused by the SZ-decrement
reduces the radio flux at 4.85 GHz by only $\leq$10\%. That is
much less than required to substantially affect 
the shape of the spectrum; for example a reduction of about 
75\% would be
required assuming the T03 data-set (Fig.~1).
Therefore, as in the previous cases,
we conclude that the steepening induced by the
SZ-decrement is negligible.

\subsection{Radio halo and SZ-correction on larger scales}

The recent WSRT observations (Brown \& Rudnick 2011)
allow the halo to be firmly traced out to unprecedentely large
scales, $\sim 0.8-0.9\, R_{500}$ radius (Fig.~3(right)).
On small apertures the flux of the halo derived from
these observations is consistent with that from Venturi et al.(1990)
(Fig.~4), but their higher sensitivity\footnote{these
observations are $\sim$4-5 times
deeper (brightness sensitivity) than those in Venturi et al.(1990)}
allows the detection of more
flux. This effect can already be seen on apertures $0.4-0.5 R_{500}$ (Fig.~1) 
and is especially strong on larger scales (Fig.~3(right)).
If we had adopted the  $\sim 0.8-0.9\, R_{500}$ scale for our calculations,
the integrated SZ decrement would
be about twice as large as that derived above;
however, this effect is more than compensated for
by the fact that the 330~MHz 
halo flux integrated on
such a large scale is also almost three times higher than
in Fig.~1 (Fig.~3, right).
Consequently, in this way we would simply re-obtain
a similar fractional decrement based on the y-radio correlation (Fig.~5).
This is shown in Fig.~6 where the observed data-points at high frequency
are compared with a bundle of power-law spectra normalized to the
330~MHz halo flux integrated on an aperture radius $=0.85\, R_{500}$
and corrected for the intervening (negative)
SZ-decrement measured by Planck on
the same aperture ($F_{obs}(\nu, \Omega_H) = 
F(\nu, \Omega_H) + \delta F_{SZ}(\nu, \Omega_H)$).
We conclude that an intrinsic power-law spectrum with a slope $\alpha
\sim 1.2-1.3$ would produce an observed flux
at 4.8 GHz that is 7-8 times higher than that measured by current 
Effelsberg observations
(T03), thus very deep single-dish observations at high frequencies
are expected to 
easily test for the presence of a spectral break in the spectrum of the 
Coma halo.

As a final remark, we note
that our approaches also assumed no influence from the SZ decrement
on scales larger than the radio halo scale.
To the extent that this is significant, it would eventually
bias low the radio ``zero'' level at high frequencies, and
consequently, 
our procedures would {\it over-estimate} the SZ-correction.
We expect, however, that 
this may affect our conclusion only at $\leq$ few percent level.

\begin{figure*}
\begin{center}
\includegraphics[width=0.95\textwidth]{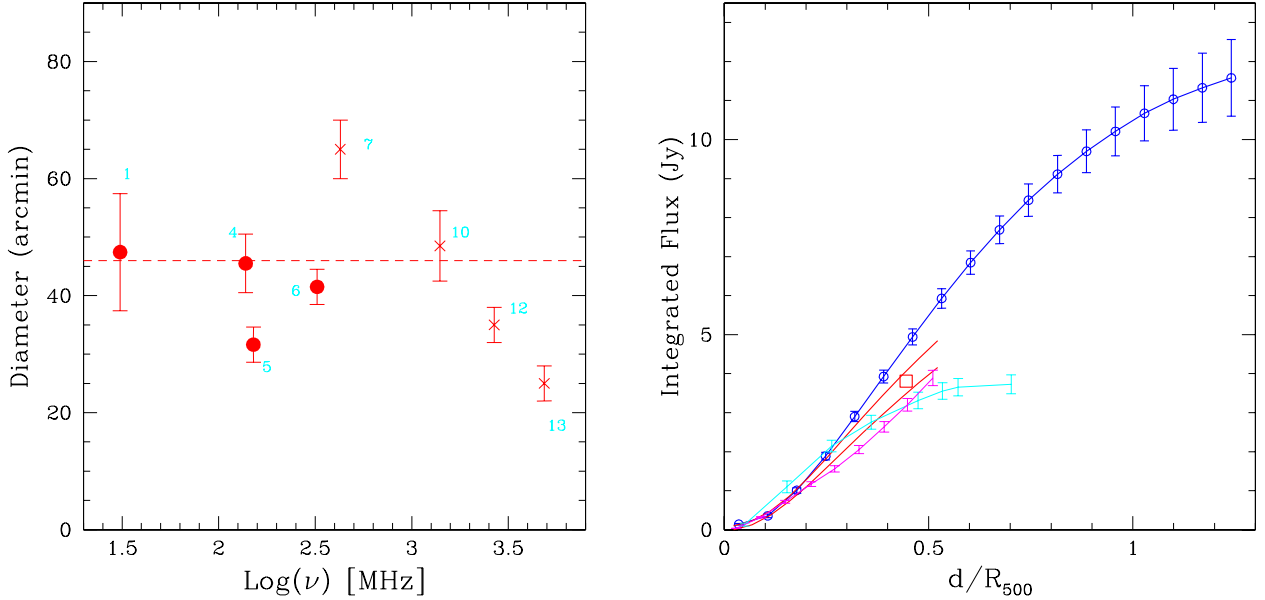}
\caption[]{
(Left): Aperture diameter used to measure the radio halo flux
as a function of the observing frequency; 
asteriscs mark single-dish observations.
The dashed line marks the representative aperture radius,
$R_H = 0.48 R_{500}$, used in our first approach to estimate
the SZ--decrement.
(Right) Full azimuthal flux profile derived from 
the Brown \& Rudnick (2011) WSRT data (solid, blue) without the quadrant
in the west, which is significantly contaminated by the tailed
radio source NGC 4869.
Errors are dominated by statistical noise at small radii 
and by the uncertainty in the ``zero level'' at large radii.
The red square marks the total flux and size of the halo from 
Venturi et al.~(1990), while the solid red lines show the integrated
flux profile obtained by Govoni et al.~(2001) rederived from these
same data.
The cyan profile is derived from Deiss et al.~(1997) at 1.4 GHz and the 
magenta profile from Pizzo (2010) at 139 MHz.
Fluxes at 139 MHz and 1400 MHz are scaled at 330 MHz using a 
spectrum $\alpha =1.22$.}
\end{center}
\end{figure*}

\begin{figure}
\begin{center}
\includegraphics[width=0.425\textwidth]{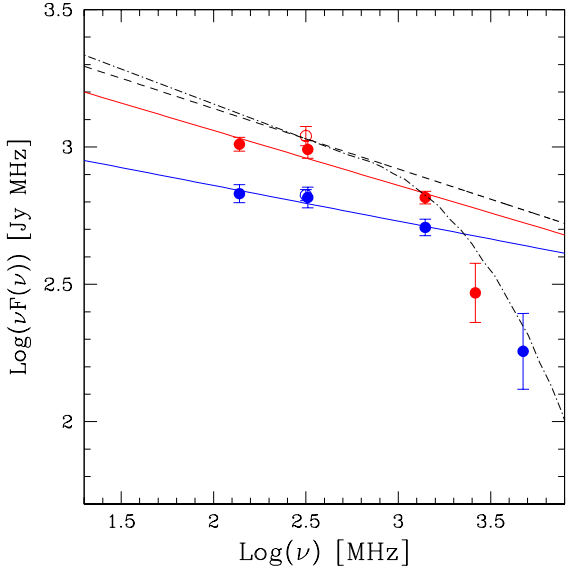}
\caption[]{
Spectrum of the radio halo extracted within 
an aperture of 17.5\arcmin (red, 0.37 $R_{500}$) 
and 13\arcmin (blue,0.27 $R_{500}$). 
The high-frequency points are corrected
for the SZ-decrement measured on the same scales. 
For comparison the empty symbols mark fluxes measured in the same
aperture radius using the Brown \& Rudnick (2011) data.
Best fits to the low-frequency data are reported as solid lines (same
color-code), while the best fit to the T03 compilation
(dashed line) and the synchrotron model with the cut-off of 
Fig.~1 (dot-dashed line) are reported for comparison.}
\end{center}
\end{figure}

\begin{figure}
\begin{center}
\includegraphics[width=0.425\textwidth]{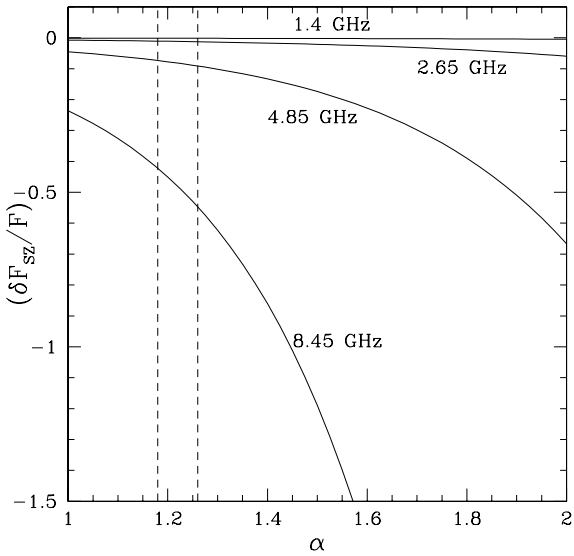}
\caption[]{
Ratio $\delta F_{SZ} / F$ as a function of the
synchrotron spectral index of the emitted spectrum
of the radio halo, assuming a power law $F(\nu) \propto \nu^{-\alpha}$
and an average halo flux in a beam area at 330 MHz $= 0.2$ Jy.
Calculations are reported at 1.4, 2.65, 4.85, and 8.45 GHz.
Vertical dashed lines mark the synchrotron spectral index of the
Coma radio halo derived from data 
at $\nu \leq 1.4$ GHz, $\alpha = 1.22 \pm 0.04$.}
\label{Fig.Lr_Lx}
\end{center}
\end{figure}

\begin{figure}
\begin{center}
\includegraphics[width=0.425\textwidth]{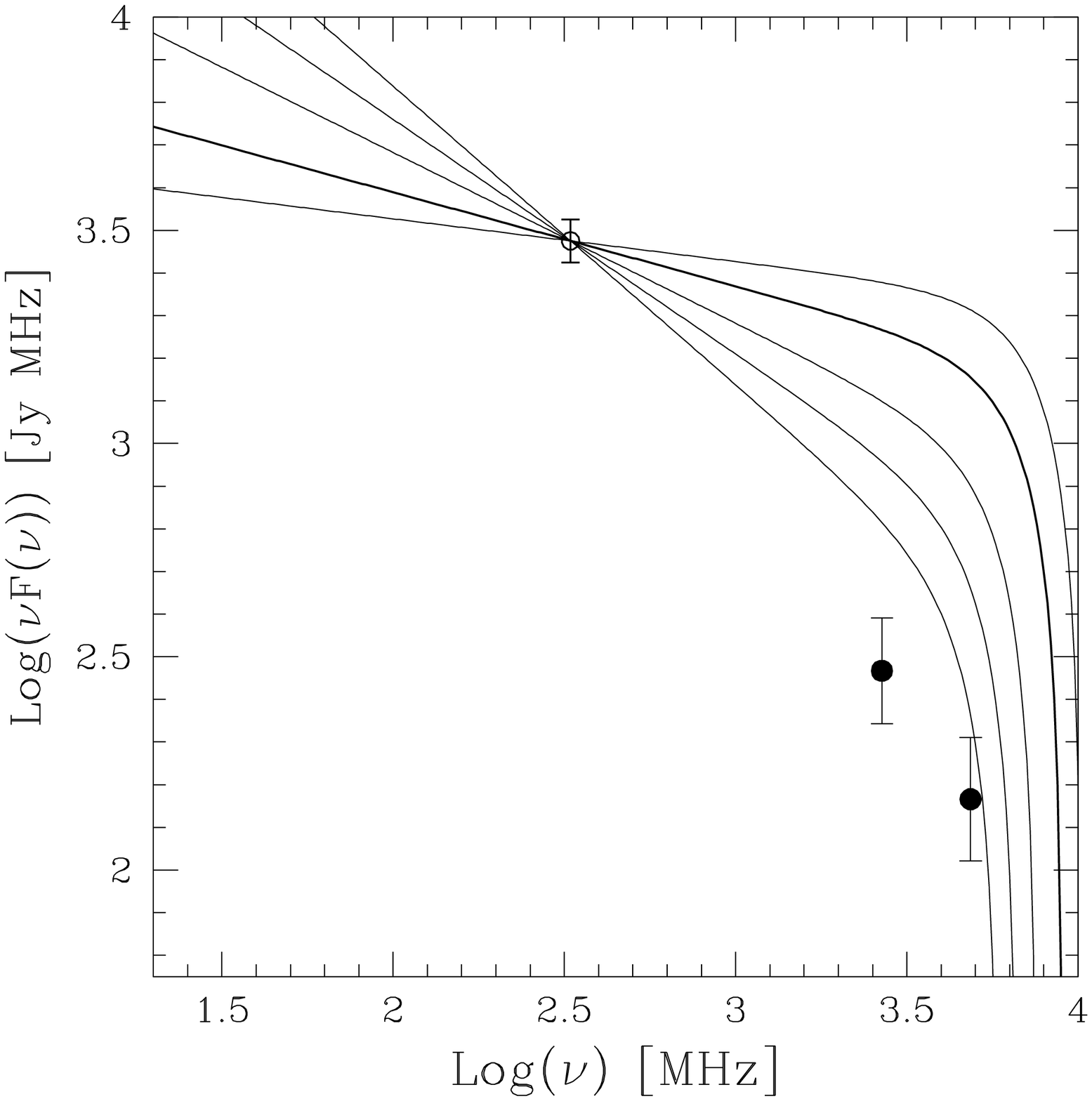}
\caption[]{Bundle of power-law spectra, $F(\nu) \propto
\nu^{-\alpha}$, 
with $\alpha=1.1$, 1.22, 1.4, 1.55, 1.7 normalized to the flux of the radio
halo derived from Brown \& Rudnick (2011) WSRT data using an aperture
radius $R_H =0.85 R_{500}$.
Models are corrected for the SZ-decrement 
($F_{obs}(\nu, \Omega_H) = 
F(\nu, \Omega_H) + \delta F_{SZ}(\nu, \Omega_H)$, Sect.~2) 
measured on the same
aperture radius. The observed high-frequency points are taken from T03.}
\end{center}
\end{figure}

\section{Conclusions}

The spectra of radio halos are important probes of the underlying
mechanisms for the acceleration of the electrons responsible for the
radio emission.
The spectrum of the Coma radio halo shows a steepening at higher
frequencies. This has triggered an on-going debate on the possibility
that this steepening is not intrinsic to the emitted radiation, but it
is caused by the intervening SZ effect with the thermal ICM.
The recent Planck data (PIPX) allow for a correct evaluation of this
effect.

\noindent
Using Planck results, 
we have shown that the negative signal caused by 
the SZ decrement does not produce a significant effect on the shape of
the spectrum of the Coma radio halo.

The spectral information of the Coma halo comes from heterogeneous
observations in the past 30 years.
For this reason, before evaluating the potential effect of the 
SZ-effect, we have discussed the main uncertainties 
on the halo spectrum that derive from the different sensitivities
of the observations at different frequencies, from the different 
apertures used to measure the flux of the
halo, and from subtracting discrete sources embedded in the halo
region.
We showed that the different sensitivities of the observation
can explain the large $\pm 30$\% scattering of the data-points observed 
in the global spectrum of the halo collected by T03.
However, we also showed that neither the different sensitivity of 
the observations (and the aperture radius of the halo), nor the
subtraction of discrete sources can naturally explain the steepening of
the halo spectrum observed at higher frequencies.

We examined the potential contribution of the SZ-effect to the
observed steepening using three complementary approaches to ensure that our
results are robust.

\noindent
With the first two methods
we measured the SZ-decrement by self-consistently adopting the aperture
radii used for flux measurements of the radio halo at the different
frequencies.
First we adopted the global compilation of data-points from T03 
and a radius $=23$ arcmin which is consistent with the aperture
used to measure the halo flux in the most sensitive observations,
between 30 MHz and 1.4 GHz.
We derived an SZ-decrement $= -1.08 (\nu/GHz)^2$mJy, which is about
four times smaller than that required to explain the observed steepening.
Second we used the available brightness profiles 
of the halo at 139, 330, and 1400 MHz to derive the spectrum of the halo 
within two fixed apertures, $=$17.5 and 13 arcmin, which correspond to the 
effective radius of the regions where the halo is detected at higher 
frequencies, 2.675 and 4.85 GHz, respectively.
In this case the flux of the halo between 139 and 1400 MHz 
is lower than that in the T03 compilation, but the SZ-signal
measured by Planck within these apertures also decreases significantly
and is about 4-5 times weaker 
than that required to explain the steepening of the spectrum measured 
within the same apertures.
As a third complementary approach we used the (almost)
scale-independent correlation between $y$ and the 330 MHz halo's flux
within a beam--aperture discovered by PIPX.
From this correlation we derived the ratio of the SZ-decrement and the radio
flux of the halo, $\delta F_{SZ}/F$, and showed that this is very low.
In particular, by assuming a spectral index of the halo
$\alpha =1.2-1.3$ the ratio is $\delta F_{SZ}/F \leq 10$\% at 4.85 GHz, 
whereas it should be $\geq 70$\% to explain the steepening observed by T03.

\noindent
Consequently, based on our analysis of the current radio data,
an intrinsic spectral break, or
cut-off, is required in the energy distribution of the electrons that 
generate the radio halo.

It is important to note, however, that the spectral analysis
presented here does not tell the whole story of Coma radio spectrum.
The recent very high sensitivity observations by
Brown \& Rudnick (2011) showed that most of the total flux
at 330 MHz is emitted beyond a radius of 20-25 arcmin, 
implying that the halo flux is significantly higher than previously thought.
The upcoming LOFAR observations
at low frequencies and more sensitive single-dish measurements 
at high frequencies have the potential of detecting the
halo on larger scales and will be essential for evaluating the global
spectral shape of the halo and possible spectral variations with radius.
For instance, we also showed that future deep observations with
single dishes at 5 GHz are expected to measure a halo flux 
on a 40 arcmin aperture radius that is predicted to
be $\sim$7-8 times higher than currently 
measured if a spectral steepening is absent,
thus providing a complementary test to our present findings.

\begin{acknowledgements}
We thank the referee for useful comments and R.Pizzo and T.Venturi
for providing useful information on their observations.
L. R. acknowledges support from the U.S. National Science Foundation, 
under grant AST-1211595 to the University of Minnesota.
J.D. acknowledges support by FP7 Marie Curie programme ``People'' of the
European Union.
K.D. acknowledges the support by the DFG Cluster of Excellence 
``Origin and Structure of the Universe''.
\end{acknowledgements}

\end{document}